\documentstyle[12pt,lathuile]{article}
\begin{document}
\title{
TOP QUARK MASS MEASUREMENTS AT THE TEVATRON
}
\author{
Gustaaf Brooijmans        \\
For the CDF and D\O \ Collaborations \\
{\em Fermi National Accelerator Laboratory, P.O. Box 500, Batavia, Illinois 60510}}
\maketitle
\baselineskip=14.5pt
\begin{abstract}
The general strategy, as well as channel-specific details, applied to the 
measurement of the top quark mass at the Tevatron in Run I are reviewed,
and the combination of the results obtained by the 
CDF and D\O \ collaborations presented.
The accelerator and detector upgrades for Run II are described, and
expected improvements in the systematic uncertainties on the measurement
evaluated.
\end{abstract}
\baselineskip=17pt
\newpage

\section{Introduction}
\label{sec:intro}
The mass of the top quark is a fundamental parameter of the standard model. 
Its precise measurement will not only improve understanding
of the model, but in combination with other experimental data it 
will also allow to further check the model's consistency.  
In addition, combined with the $W$ mass measurement, it
can be used to constrain the mass of the Higgs boson.  

This paper is organized 
as follows: Section~\ref{sec:prod} addresses top quark production and
decay topologies at the Tevatron;  top-quark mass measurements performed
in Run I are reviewed in Section~\ref{sec:measure}, and the combined 
CDF and D\O \ result is given in Section~\ref{sec:combine}.  This is followed
in Section~\ref{sec:upgrade} by a description of accelerator and 
detector upgrades for Run II, and in Section~\ref{sec:system} by the prospects
 for reduction of systematic 
uncertainties on the mass measurement.
Conclusions are drawn in Section~\ref{sec:conclude}.

\section{Top Quark Production and Decay at the Tevatron}
\label{sec:prod}
At the Tevatron, top quarks are produced mainly in pairs.  At 
$\sqrt{s} = 1.8\ TeV$, about 90\% of the cross section is due to 
$q \overline{q}$ annihilation and 10\% to gluon fusion.  The 
$t \overline{t}$ production cross section has been measured by both
the CDF\cite{cdfxsec} and D\O\cite{d0xsec}\ collaborations to be
$\sigma(m_{top} = 175\ GeV/c^2) = 6.5^{+1.7}_{-1.4}\ pb$, and 
$\sigma(m_{top} = 172\ GeV/c^2) = 5.9 \pm 1.7\ pb$, respectively.  
These results
are in good agreement with the theoretical predictions\cite{theoxsec}, 
which range from $4.7$ to $5.5\ pb$.
Single top quarks can also be produced in electroweak processes, but 
due the smaller cross sections and larger backgrounds, these have not
yet been observed.

Because of its large width, the top quark decays to a $b$ quark and a
$W$ boson before it hadronizes.  The various possible $W$ boson decays 
determine the separate channels for mass measurement: events in which both
$W$ bosons decay hadronically comprise the ``all hadronic'' channel, 
and account
for $\approx$ 44~\% of all produced $t \overline{t}$\ events. 
Events in which one or both of the $W$ bosons decay to $e\nu$ or $\mu\nu$
form the ``lepton+jets'' ($\approx$ 30~\%) and ``dilepton'' ($\approx$ 4.9~\%) 
channels, respectively.
Events in which one or both $W$ bosons decay to $\tau\nu$ are not used
in the mass and cross section measurements.

\section{Top-Quark Mass Measurements}
\label{sec:measure}

\subsection{General Strategy}
\label{subsec:strat}

While different experiments and different channels require specific 
analyses, these measurements have enough in common so that one can speak 
of a general strategy 
to extract the mass of the top quark from data.  This involves four steps:
\begin{itemize}
\item Apply a set of criteria to select the candidate sample, 
while being careful
to choose cuts that introduce little bias in the extracted mass.
\item Fit each event's kinematics to the top quark pair production hypothesis,
extracting a fitted top-quark mass $m_{fit}$.
\item Signal distributions
are generated for many values of $m_{top}$, and 
a likelihood (${\mathcal L}$) fit is used to compare the measured $m_{fit}$ 
distribution to 
summed signal and background distributions for each value of $m_{top}$.
The fit determines the 
relative contributions of signal and background.
\item Fit the obtained $ln {\mathcal L}$ versus $m_{top}$,
and extract the mass at the minimum of $-ln {\mathcal L}$.
\end{itemize}

In all cases, $t \overline{t}$ signal distributions have been generated 
using the 
HERWIG\cite{herwig} Monte Carlo generator.  The background is dominated by
multiple-jet production and the production
of a $W$ boson in association with additional jets.  The former is derived
from data while the latter is modeled using the 
VECBOS\cite{vecbos} event generator.

\subsection{All-Hadronic Channel}
\label{subsec:allhad}

Only CDF has published a mass measurement\cite{CDFallhad} in the 
all-hadronic channel.
Events are required to have six or more jets, of which at least one
is tagged as a $b$-quark jet by the detection of a secondary vertex 
(``SVX tag'').   Imposing that the masses of
the top and anti-top quark be equal, and that the invariant mass
of pairs of jets forming $W$ bosons equals the $W$-boson mass, yields 
a 3-C fit.  However, for each event, multiple combinations are possible in
the association of jets to $W$ bosons or $b$ quarks, so  
the combination resulting in the lowest $\chi^2$ is chosen
as the right kinematic fit.
The sample consists of 136 events (corresponding to an integrated luminosity
of $109\ pb^{-1}$), with the background contamination estimated at 
108$\pm$9~events.  The measured top mass is 
$m(top) = 186.0 \pm 10\ (stat) \pm 5.7\ (syst)\ GeV/c^2$.

\subsection{Lepton + Jets Channel}
\label{subsec:l+jets}

The D\O \ lepton + jets analysis\cite{D0l+jets} requires an isolated 
central lepton, four or more jets and at least $20\ GeV$ missing transverse 
energy.  Since the neutrino's
longitudinal momentum is not measured, the kinematic fit to the 
$t \overline{t}$ hypothesis is now a 
2C fit.  D\O \ also computes a discriminant {\bf D}, based on 4 variables
weakly correlated with the top-quark mass.  This discriminant is computed
through a comparison of signal and background using classical methods, as
well as a neural network.  The likelihood fit
to determine the top mass is then performed on a two-dimensional grid of
{\bf D} versus the top mass.  Both discriminants yield results that are 
in good agreement with each other, which when combined give
$m(top) = 173.3 \pm 5.6\ (stat) \pm 5.5\ (syst)\ GeV/c^2$.  The background 
contamination is estimated to be approximatively 50 out of the 77 events 
that pass all selection criteria.

\begin{figure}[t]
 \vspace{8.5cm}
\includegraphics{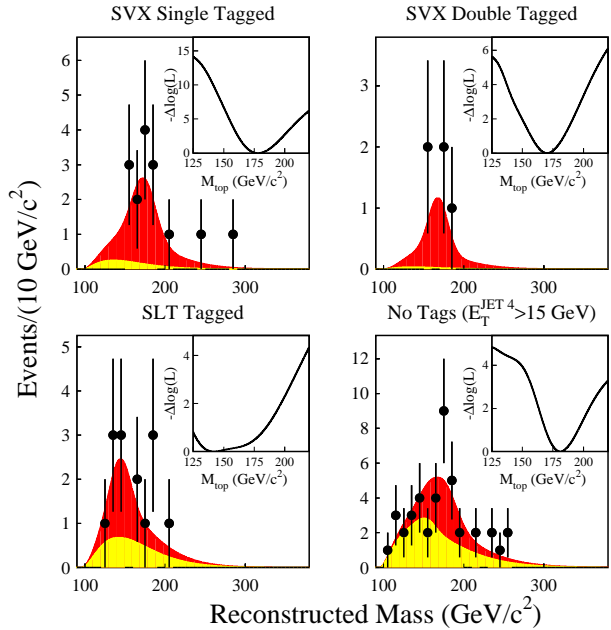}
 \caption{\label{fig:CDFl+jets} \it  CDF lepton  + jets analysis: reconstructed mass
     distributions in each of the four subsamples (see text).  Each plot shows the data 
     (points), the result of the combined fit to top + background (dark shading), 
     and the background component of the fit (light shading).  The insets show the 
     variation of the log-likelihoods with $m_{top}$ for the separate fits.}
\end{figure}

CDF\cite{CDFl+jets} uses a similar procedure, but finds that 
optimal results can be obtained by dividing the sample into 4 subsamples: 
events
with two SVX tags (5 events), events with a single SVX tag (15 events), events
without SVX tags but where one jet is identified as a $b$-quark jet through 
a soft lepton tag (``SLT tag'') (14 events), and finally events without any tags
(42 events).  The samples are fitted separately and the results, shown in 
Fig.~\ref{fig:CDFl+jets}, when combined yield
$m(top) = 175.9 \pm 5.1\ (stat) \pm 5.3\ (syst)\ GeV/c^2$.

\subsection{Dilepton Channel}
\label{subsec:dilept}

As indicated in Section~\ref{sec:prod}, the dilepton channel has 
fairly low statistics.  On the other hand, the background is small, 
dominated by $Z$ boson decays to charged leptons and particle 
misidentification.
The difficulty arises from the presence of two neutrinos, for which only the 
combined transverse momentum can be measured.  The system is thus kinematically
underconstrained, and both experiments use the so-called ``neutrino weighting''
technique to measure the top quark mass: for every chosen top quark mass, 
there is a sampling of allowed 
neutrino rapidities.  Given the neutrino rapidities, the assumed top
mass, and the charged-lepton and b-quark momenta, the system can
now be solved for the transverse and longitudinal momentum components 
of the neutrinos (with a two-fold ambiguity for each neutrino).  
A weight is then assigned to each solution based on the agreement
between the calculated and measured missing transverse momentum.  All weights
are summed for all possible combinations of neutrino rapidities.
This is done as a function of top mass value for each event.

\begin{figure}[t]
 \vspace{8.2cm}
\includegraphics{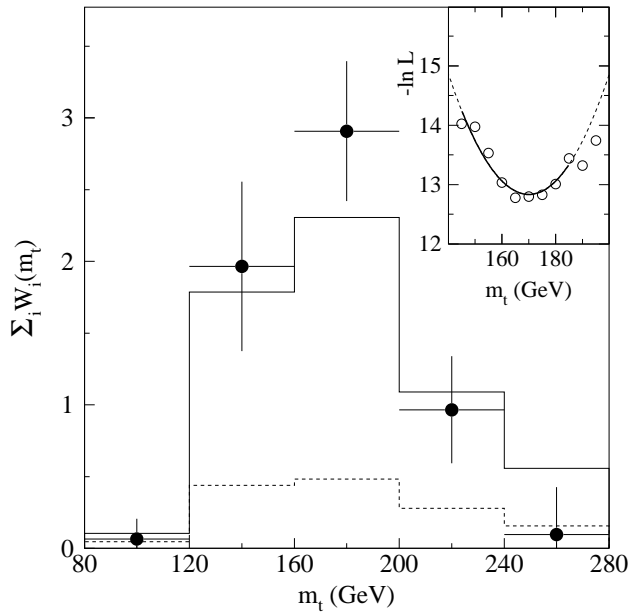}
 \caption{\label{fig:D0dilept} \it  D\O \ dilepton analysis: summed event weight 
   function $\sum_i W_i$ for the data sample (points), the sum of
   the fitted signal and background (solid), and the background alone (dashed).
     The error bars indicate the rms observed for five-event samples in 
     ensemble tests.  The inset shows the corresponding fit to $-ln {\mathcal L}$,
     drawn as a solid curve in the region considered in the fit.}
\end{figure}

D\O\cite{D0dilept} has observed 6 events in a sample corresponding to an 
integrated luminosity of about $125\ pb^{-1}$.  For each event, the 
distribution of weights is split into five mass regions, and used to form 
a vector for calculating the mean mass value for all events in the sample.
Fig.~\ref{fig:D0dilept}
summarizes the results based on these events.
D\O \ also evaluates the mass in this sample using an extended version of 
the Dalitz-Goldstein and Kondo method.  Both results are compatible,
and the combined measured top-quark mass is
$m(top) = 168.4 \pm 12.3\ (stat) \pm 3.6\ (syst)\ GeV/c^2$.

Eight events are observed by CDF\cite{CDFdilept} (in a sample corresponding 
to an integrated luminosity of $109\ pb^{-1}$).  Each event's top mass
estimate is determined from the peak of the weight distribution, and
the resulting distribution compared with Monte Carlo simulation,
leading to a measured top-quark mass of
$m(top) = 167.4 \pm 10.3\ (stat) \pm 4.8\ (syst)\ GeV/c^2$.

\subsection{Systematic Uncertainties}
\label{subsec:system}

The main contributions to the systematic uncertainties on the Run I
top-quark mass measurements are given in Table~\ref{tab:system}.
The dominant contribution is due to the determination
of the jet energy scale.  Other important contributions come
from uncertainties in initial and final state radiation, and modeling
of the background.  
All of these will be discussed
in more detail in Section~\ref{sec:system}.

\begin{table}
\centering
\caption{\label{tab:system} \it Summary of main contributions  to 
systematic uncertainties (in $GeV/c^2$) in the measurement of the mass of the 
top quark in Run I:  a) jet energy scale, b) model for signal (mainly initial 
and final-state radiation), c) Monte Carlo generator, d) Uranium noise
and multiple interactions, e) model for background, and f) method for 
mass fitting.
Contributions to the uncertainties
differ somewhat between the two experiments.  For example, for CDF 
measurements, the uncertainty
due to multiple interactions is included in the uncertainty in the 
jet energy scale.}
\vskip 0.1 in
\begin{tabular}{|l|c|c|c|c|c|c|} \hline
          &  a & b & c & d & e & f \\
\hline
\hline
CDF Dilepton      & 3.8 & 2.8 & 0.6 & - & 0.3 & 0.7 \\
CDF Lepton + Jets & 4.4 & 2.6 & 0.1 & - & 1.3 & - \\
CDF All Hadronic  & 5.0 & 1.8 & 0.8 & - & 1.7 & 0.6 \\
D\O \ Dilepton      & 2.4 & 1.7 & - & 1.3 & 1.0 & 1.1 \\
D\O \ Lepton + Jets & 4.0 & 1.9 & - & 1.3 & 2.5 & 1.5 \\
\hline 
\end{tabular}
\end{table}

\section{Combined CDF and D\O \ Measurement}
\label{sec:combine}

\begin{figure}[t]
 \vspace{8.7cm}
\includegraphics{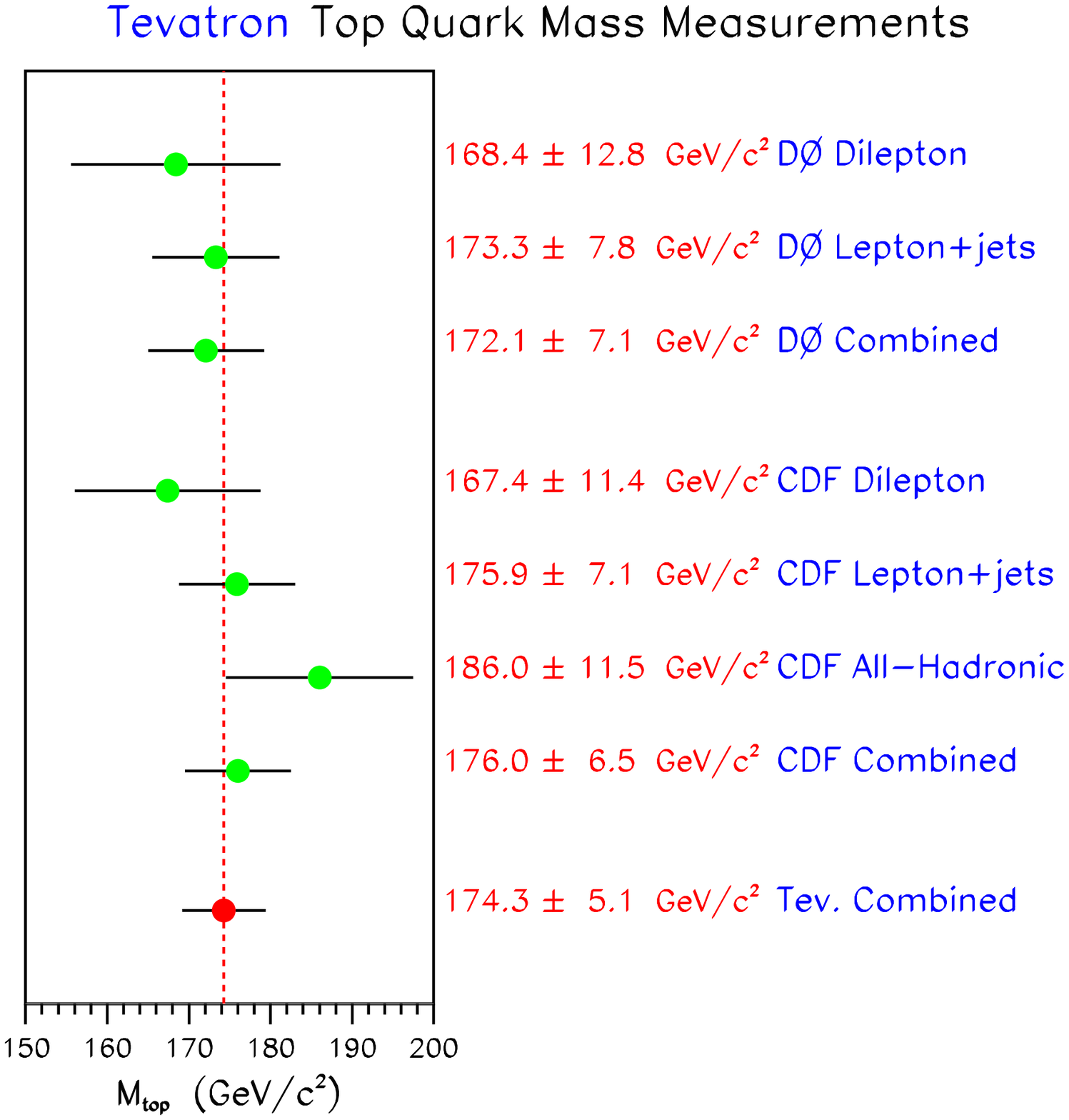}
 \caption{\label{fig:topm-comb} \it Individual and combined top-quark 
mass measurements at the Tevatron.}
\end{figure}

The Top Averaging Group\cite{Tevcomb} has combined the 5 mass
measurements described in Section~\ref{sec:measure}.  As expected, after 
taking into account all correlations, the best contribution comes from the 
lepton + jets channel.  The combined result is
$m(top) = 174.3 \pm 3.2\ (stat) \pm 4.0\ (syst)\ GeV/c^2$, which is 
compared with the individual measurements in Fig.~\ref{fig:topm-comb}.

\section{Run II Upgrades}
\label{sec:upgrade}

Run II is formally scheduled to start at the Tevatron on March 1st, 2001.  
This 
will follow a shutdown during which both the accelerator complex and 
the collider detectors will have undergone significant upgrade from their
Run I configurations.

\subsection{Upgrade of the Accelerator Complex}
\label{subsec:accupgr}

In Run II, the Tevatron is expected to operate at a center of mass energy
$\sqrt{s} = 2 \ TeV$, an increase of $200 \ GeV$ over Run I.  For top quark
studies, this yields a substantial rise of 40 \% in the $t \overline{t}$ 
cross-section.

In addition to that, the completion of the Main Injector should provide 
significant improvement in the luminosity.  Both experiments are expected
to record data corresponding to an integrated luminosity of $2\ fb^{-1}$ in
the first two years, followed, after a brief shutdown, by $5\ fb^{-1}$ 
per year until the start up of the LHC.

\subsection{D\O \ Upgrade}
\label{subsec:d0upgr}

The main features of the D\O \ upgrade involve the addition of a 2 Tesla 
solenoid inside the calorimeter, combined with improved tracking and
particle identification capabilities.  

Two central tracking systems have
been built: an extensive silicon detector consisting of barrels for
central tracking and disks covering the forward regions, and a 
scintillating fiber tracker between the silicon and the solenoid.
The silicon barrels
have four layers mainly of double-sided silicon sensors, while all
disks provide two hit projections.  
The fiber tracker consists of 8 coaxial cylinders, and is used in the 
Level-1 trigger.

In addition to the tracking systems, preshower detectors will be used 
to identify 
electrons and photons.  Also, the muon coverage has been extended and a new
three-level trigger system is being built.

\subsection{CDF Upgrade}
\label{subsec:cdfupgr}

The upgraded CDF detector will have improved central tracking 
detectors, as well as increased coverage in pseudorapidity for lepton
identification.

The new silicon detector consists of three parts: ``Layer-00'', 
silicon barrels, and the ``Intermediate Silicon Layers (ISL)''.  Layer-00
is a single layer of silicon located very close to the beam, leading
to significant improvement in impact parameter resolution.  The barrels
have five layers of double-sided silicon sensors, with the ISL located 
at larger radii, providing additional points between the silicon and the
Central Outer Tracker (COT).  The COT is an improved version of CDF's 
Run I central drift chamber: it has 12 layers of wires, less material, and
operates with shorter drift times.

New end plug calorimeters have been built, giving better coverage for 
electron identification, and the muon coverage has also been increased.
In addition to this, a time-of-flight system is located between the 
COT and the solenoid, and a new three-level trigger system is being
implemented.

\section{Systematic Uncertainties on the Top-Quark Mass Expected in Run II}
\label{sec:system}

In this section, we discuss ways to reduce the dominant systematic 
uncertainties on the mesurement of the mass of
the top quark.

\subsection{Determination of Jet Energy Scale}
\label{subsec:jetscale}

For Run I data, CDF and D\O\ use a somewhat different approach to 
determine the jet energy scale.
In D\O, the essential steps of the procedure are the following: the
electromagnetic scale is normalized by the mass of the $Z$-boson, 
and the hadronic
response is then derived by balancing $\gamma$ + jets events.  Corrections
for the underlying event, multiple interactions and noise, are determined
from minimum-bias data, while corrections for leakage outside the jet
cone are computed based on test-beam data and Monte Carlo simulation.
Finally, the conversion from jet to parton energy is based on Monte Carlo
(HERWIG), with separate treatment of light and $b$-quark jets.
A small remnant pseudo-rapidity dependence is corrected using data from 
$\gamma$ +  1 jet events, and the results are cross-checked
using $(Z \rightarrow ee)$ + jets events.
In D\O , the required corrections are determined separately and 
symmetrically for data and Monte Carlo, and then the relative difference
between data and Monte Carlo is added to the systematic uncertainty.  
Since the top-quark mass is
measured by comparison of data with simulation, this is a valid approach.

While many of the steps are similar in CDF, there is one fundamental
difference: in CDF, the Monte Carlo simulation, and in particular the
fragmentation, is tuned to the data, exploiting the available accurate 
tracking information.
As is shown in Table~\ref{tab:system}, the two different approaches do not
seem to lead to significantly different results.

In Run II, the additional and improved data will yield at least two new
handles on the jet energy scale in $t\overline{t}$ events.  The first is 
given by the larger sample of lepton + jets events with two $b$-quark
tags.  In 2 $fb^{-1}$, this sample is expected to contain of the order
of 300 to 500 events, allowing accurate calibration of light quark jets
through the reconstruction of the $W$-boson mass.  
Probably the most precise way to determine the calibration of the 
jet energy scale is given 
by the large sample of $Z \rightarrow b \overline{b}$ events expected in 
Run II.  This type of events has already been observed by CDF in 
Run I\cite{zbb},
but the sample is unfortunately to small to be used for determining the
jet energy scale.  It is expected that using these events in Run II will
allow reduction in this dominant uncertainty to below 1\%.

\subsection{Initial and Final-State Radiation}
\label{subsec:fsr}

A second sizeable contribution to systematic uncertainty is due to the
treatment of initial and final-state radiation.  Clearly,
additional jets from initial-state radiation should not be
included in the kinematic fits, but those due to final-state radiation 
certainly should be.
Unfortunately, such additional jets are the least energetic only about 
55 \% of the time.
However, theoretical work is progressing, and present and future versions
of the Monte Carlo generators should reproduce these processes with better
accuracy.
Furthermore, the sample of lepton + jets events with extra jets in Run II
should be large enough to allow more detailed study and comparison to 
Monte Carlo.  This can be done using variables such as the number of 
additional jets, relative angles between jets, etc.

\subsection{Backgrounds}
\label{subsec:backgr}

Multiple-jet QCD background is determined directly from data, and does not
represent a major problem.  In Run I, a significant systematic uncertainty
came from the $W$+jets background, which was modeled using the VECBOS event
generator.  In the meantime, $W$ and $Z$-boson transverse momentum 
distributions at the Tevatron have been studied in detail\cite{zpt}, 
contributing
to our understanding of these vector boson production processes.  Better
Monte Carlo generators already exist, so this uncertainty should be 
reduced appreciably in Run II.

\section{Conclusions}
\label{sec:conclude}

In Run I at the Tevatron, the top quark mass has been measured with an 
accuracy of 3 \%.  The measured value is 
$m(top) = 174.3 \pm 3.2\ (stat) \pm 4.0\ (syst)\ GeV/c^2$, which corresponds
to a top Yukawa coupling 
$\lambda_{top}(Q^2 = m_{top}^{2}) = 1.00 \pm 0.03$.

In Run II, thanks to the improved accelerators and detectors, statistics 
will not be an issue.  Furthermore, the additional data will provide better
control of systematic uncertainties, where the main contribution will
probably shift to the treatment of initial and final-state radiation.  A
challenging goal will be to reduce the systematic uncertainty on the top 
quark mass to less than 2 $GeV/c^2$.

\end{document}